\documentstyle{article}

\newcommand{\bsigma}{\mbox{{\boldmath $\sigma$}}}
\newcommand{\blambda}{\mbox{{\boldmath $\lambda$}}}

\makeatletter
\@addtoreset{equation}{section}
\makeatother

\begin{document}
\begin{flushright} NUP-A-98-5\\hep-th/9801164\\(revision 2) \end{flushright}
\vspace{0.7cm}
\begin{center}
{\large
Constraints among Coupling Constants in Noncommutative Geometry Models}\\
\vspace{0.7cm}
Eizou UMEZAWA\footnote{e-mail: umezawa@phys.cst.nihon-u.ac.jp}\\
\vspace{0.2cm}
{\it Department of Physics, College of Science and Technology,\\
Nihon University, Tokyo 101-8308}
\end{center}
\vspace{0.1cm}
\begin{abstract}
We study constraints among coupling constants of the standard model obtained in the noncommutative geometry (NCG) method. First, we analyze the evolution of the Higgs boson mass under the renormalization group by adopting the idea of \'Alvarez et al. For this analysis we derive two certain constraints by modifying Connes's way of constructing the standard model. Next, we find renormalization group invariant (RGI) constraints in the NCG method. We also consider the relation between the condition that a constraint among coupling constants of a model becomes RGI and the condition that the model becomes multiplicative renormalizable by using a simple example.
\end{abstract}
\section{Introduction}
The noncommutative geometry (NCG) method proposed by Connes and Lott points out a clear geometric significance to the Higgs sector of the standard model.\cite{1,2,3,4,5} \ It has been noted that this method gives rise to certain constraints among coupling constants of the model, which imply some physical predictions such as mass relations among the Higgs boson, the top quark, and the $W$ boson.\cite{5,6,7,MP} \ However, the multiplicative renormalizability of the model obtained from this method is a question, because the constraints do not seem to be consequences of any symmetry, as far as we know at this time.

As a way of imposing constraints among coupling constants of a given model in quantum field theory, the renormalization group invariant (RGI) method is known.\cite{8,9} \ \'Alvarez et al. showed that the NCG constraints of the standard model given in Ref.~\cite{2} are not invariant under the renormalization group (RG) evolution.\cite{10,11} \ They adopted the point of view that the constraints hold only at a given energy scale, and they then analyzed the RG evolution of the Higgs and the top mass by using the constraints.\cite{11} \ Subsequently, several authors performed renormalization group analyses of the Higgs boson mass \cite{12,13} \ by using the constraints coming from their own methods that have a close relation to the Connes-Lott method.\cite{14}

In this paper we begin with a review of the construction of the standard model in the NCG method and make clear the derivation of the constraints among coupling constants coming from this method. We find that there is an arbitrariness in the construction of the bosonic lagrangian. Constraints are obtained by imposing some restrictions on the way of making the bosonic lagrangian from the field strengths defined in NCG. Based on this observation, we make the following two considerations.

First we obtain certain constraints in a natural manner, and study the RG evolution of the Higgs boson mass by adopting the idea of \'Alvarez et al. In Ref.~\cite{2}, Connes gives two relations among coupling constants, one of which contains the quartic self-coupling constant of the Higgs field. However, there is only a single essential constraint, because the relations contain one free parameter. To predict the Higgs boson mass from the NCG constraints, we need two constraints: one of these is used to determine the energy scale at which the NCG constraints hold, and the other is used to give a initial condition to solve the differential equation that determine the RG evolution of the quartic self-coupling constant of the Higgs field. We give two constraints by modifying Connes's construction of the standard model.

Second, we consider RGI constraints in the NCG method. We can give even the most general lagrangian of the standard model by using the arbitrariness in the manner of constructing the bosonic lagrangian. In this sense, we can say that the NCG method gives a parametrization of the coupling constants of the standard model different from the one usually used. It is interesting to consider how the NCG parametrization works to give RGI constraints of the standard model.

Even if we give a model whose coupling constants obey the RGI constraints, the multiplicative renormalizability of the model is still an open question. In this paper we also consider the relation between the condition that a constraint among coupling constants of a model becomes RGI and a condition required by the multiplicative renormalizability of the model whose bare coupling constants are constrained by the same constraint.

In the next section, we review the standard model of Connes and Lott. We generalize the manner of constructing the bosonic lagrangian and consider the constraints coming from the NCG method. In \S 3, we adopt the idea of \'Alvarez et al., and study the RG evolution of the Higgs boson mass by using constraints that are obtained by imposing a natural restriction on the way of constructing the bosonic lagrangian. In \S 4, we consider RGI constraints of the standard model in the NCG method. In \S 5, we discuss the relation between the condition that a constraint among coupling constants of a model becomes RGI and the condition that the model becomes multiplicative renormalizable by using a simple example. Section 6 is devoted to conclusion and discussion.
\section{The standard model from the NCG method and constraints among coupling constants}
In the NCG method, the action of the standard model is constructed as a Yang-Mills action on a product space of the usual 4-dimensional continuum $M$ by a finite space $F$. The geometry of $M \times F$ is defined by Connes's NCG. The basic data of NCG is a triplet $(A, {\cal H}, D)$. 

First, $A$ is an involutive algebra. For the ordinary manifold $M$, this is the algebra of smooth functions on $M$. The exterior differential $d$ on elements of $A$ is defined as an operation satisfying
\begin{equation}
d(a_{1}a_{2})=da_{1}\cdot a_{2}+a_{1}da_{2},~~a_{1,2} \in A.
\end{equation}
Elements of the space of all differential $k$-forms $\Omega ^{(k)}(A)~ (k=0,1, \cdots)$ have the form
\begin{equation}
a_{1}da_{2}\cdots da_{k},~~a_{1}, \cdots a_{n} \in A,
\end{equation}
and the operation $d$ on it is defined by
\begin{equation}
d(a_{1}da_{2}\cdots da_{k})=da_{1}da_{2}\cdots da_{k}.
\end{equation}

Second, $\cal{H}$ is a Hilbert space. Elements of $A$ are represented as operators on $\cal{H}$. We write a representation of $a \in A$ as $\pi(a)$.

Third, $D$ is a self-adjoint operator on $\cal{H}$ that is used to represent the differential forms on $\cal{H}$:
\begin{equation}
\pi(da)=[D,\pi(a)],~~a\in A.
\end{equation}

For $M \times F$, the triplet is defined as follows.
\begin{itemize}
  \item $A$ : Two algebras,
\begin{eqnarray}
{\cal A}&=&C^{\infty}(M)\otimes ({\bf C}\oplus {\bf H}),\label{A}\\
{\cal B}&=&C^{\infty}(M)\otimes ({\bf C}\oplus M_{3}({\bf C}))\label{B},
\end{eqnarray}
are introduced, where $C^{\infty}(M)$ is the algebra of smooth functions on $M$,\footnote{$M$ is a 4-dimensional Euclidean space, so we use the metric $g^{\mu \nu}=-\delta^{\mu \nu}~(\mu ,\nu =1 \sim 4)$ and gamma matrices satisfying $\{\gamma^{\mu}, \gamma^{\nu}\}=-2\delta^{\mu \nu}$ and $\gamma^{\mu \dag}=-\gamma^{\mu}$. We will obtain a Euclidean version of the standard model. We return it to the Minkowski version after construction of the lagrangian.} \ ${\bf C}$ is the space of all complex numbers, ${\bf H}$ is the quaternion algebra, and $M_{3}({\bf C})$ is the algebra of $3 \times 3$ matrices.

  \item ${\cal H}$ : Both ${\cal A}$ and ${\cal B}$ are represented in a common space ${\cal H}$. This space is a direct sum of the leptonic and the quark parts: ${\cal H}={\cal H}_{l}\oplus {\cal H}_{q}$, where ${\cal H}_{l}$ is the space of 4-spinor-valued 3$N$ $\times $1 vectors, and ${\cal H}_{q}$ is the space of 4-spinor-valued $4\cdot 3 N\times 1$ vectors. Here, $N$ corresponds to the generation number of fermions. Representations of $a \in {\cal A}$ and $b \in {\cal B}$ have the forms
\begin{equation}
\pi (a)=\pmatrix{\pi_{l}(a)&0\cr 0&\pi_{q}(a)\cr },\label{component1}
\end{equation}
where
\begin{eqnarray}
\pi_{l}(a)&=&\pmatrix{f&0&0\cr 0&\alpha &\beta \cr 0&-\beta^{\ast}&\alpha^{\ast}\cr}\otimes 1_{N}\otimes 1_{C},\\
\pi_{q}(a)&=&\pmatrix{f&0&0&0\cr 0&f^{\ast}&0&0\cr 0&0&\alpha &\beta \cr 0&0&-\beta^{\ast}&\alpha^{\ast}\cr}\otimes {\bf 1}_{3}\otimes 1_{N}\otimes 1_{C},
\end{eqnarray}
and
\begin{equation}
\pi (b)=\pmatrix{\pi_{l}(b)&0\cr 0&\pi_{q}(b)\cr },\label{component2}
\end{equation}
where
\begin{eqnarray}
\pi_{l}(b)&=&g1_{l}\otimes 1_{N}\otimes 1_{C},\\
\pi_{q}(b)&=&1_{q}\otimes {\bf g}\otimes 1_{N}\otimes 1_{C},
\end{eqnarray}
respectively. Here,
\begin{equation}
f\in {\bf C},~~\pmatrix{\alpha &\beta \cr -\beta^{\ast}&\alpha^{\ast}\cr} \in {\bf H},~~g\in {\bf C},~~{\bf g}\in M_{3}({\bf C}),
\end{equation}
and $1_{N}$, $1_{C}$ and ${\bf 1}_{3}$ are the respective units of $N \times N$ matrices, the Clifford algebra of the gamma matrices and $M_{3}({\bf C})$, and
\begin{equation}
1_{l}=\pmatrix{1&0&0\cr 0&1&0\cr 0&0&1\cr},~~1_{q}=\pmatrix{1&0&0&0\cr 0&1&0&0\cr 0&0&1&0\cr 0&0&0&1\cr }.
\end{equation}
All matrix elements of $\pi(a)$ and $\pi(b)$ are also functions on $M$.
  \item $D$ : 
\begin{eqnarray}
D&=&D^{M}+D^{F},\\
D^{M}&=&\partial_{\mu}\pmatrix{1_{l}\otimes 1_{N}&0\cr 0&1_{q}\otimes {\bf 1}_{3}\otimes 1_{N}}\otimes \gamma^{\mu},\\
D^{F}&=&\pmatrix{D_{l}^{F}&0\cr 0&D_{q}^{F}\otimes {\bf 1}_{3}\cr}\otimes \gamma^{5},
\end{eqnarray}
where
\begin{equation}
D_{l}^{F}=\pmatrix{0&M_{e}^{\dag}&0\cr M_{e}&0&0\cr 0&0&0\cr},~~D_{q}^{F}=\pmatrix{0&0&M_{d}^{\dag}&0\cr 0&0&0&M_{u}^{\dag}\cr M_{d}&0&0&0\cr 0&M_{u}&0&0\cr}.
\end{equation}
Here, $M_{e,u,d}$ and $0$ contained in $D_{l,q}^{F}$ are $N \times N$ matrices.
\end{itemize}

For later use, we define the following decomposition of the differential $d$:
\begin{equation}
d=d^{M}+d^{F}
\end{equation}
with
\begin{eqnarray}
d^{A}(a_{1}a_{2})&=&(d^{A}a_{1})a_{2}+a_{1}d^{A}a_{2},\\
d^{A}(a_{1}d^{B}a_{2} \cdots d^{C}a_{n})&=&d^{A}a_{1}d^{B}a_{2} \cdots d^{C}a_{n},~~a_{1\sim n} \in {\cal A}\mbox{ or }{\cal B},
\end{eqnarray}
where $A,B,C=M\mbox{ or }F$. Representations of $d^{M}a$ and $d^{F}a$ ($a \in {\cal A}\mbox{ or }{\cal B}$) are defined by
\begin{equation}
\pi(d^{M}a)=[D^{M},\pi(a)],~~\pi(d^{F}a)=[D^{F},\pi(a)],
\end{equation}
respectively. We note that $\pi(da)=\pi(d^{M}a)+\pi(d^{F}a)$. This decomposition of the differential is not only practically convenient for the calculation but also important to consider a generalization of the way to make gauge and Lorentz invariants in this method.

We can consider a gauge theory on $M \times F$. The vector spaces ${\cal E}_{\cal A}$ and ${\cal E}_{\cal B}$ of all sections of vector bundles associated with ${\cal A}$ and ${\cal B}$ are defined as
\begin{equation}
{\cal E}_{\cal A}={\cal A},~~{\cal E}_{\cal B}={\cal B},
\end{equation}
respectively. We define the actions of ${\cal A}$ on ${\cal E}_{\cal A}$ and ${\cal B}$ on ${\cal E}_{\cal B}$ to both be left actions for convenience. Fermion fields belong to a quotient set of $\pi({\cal E}_{\cal A})\otimes \pi({\cal E}_{\cal B})\otimes {\cal H}$ by the equivalence relation
\begin{equation}
\pi(a\xi_{\cal A})\otimes \pi(b\xi_{\cal B})\otimes \eta \sim \pi(\xi_{\cal A})\otimes \pi(\xi_{\cal B})\otimes \pi(a)\pi(b)\eta
\end{equation}
for all $a \in {\cal A},~b \in {\cal B},~\xi_{\cal A} \in {\cal E}_{\cal A},~\xi_{\cal B} \in {\cal E}_{\cal B}$ and $\eta \in {\cal H}$. The fields are expressed as
\begin{equation}
\Psi=(e_{R},e_{L},\nu_{L},d_{R},u_{R},d_{L},u_{L})^{T},
\end{equation}
where the components of $\Psi$ are $N \times 1$ vectors of the generation, and $u_{R,L}$ and $d_{R,L}$ are $SU(3)_{c}$ triplets.

The gauge groups $U_{\cal A}$ and $U_{\cal B}$ on ${\cal E}_{\cal A}$ and ${\cal E}_{\cal B}$ are defined by
\begin{eqnarray}
U_{\cal A}&=&\{u_{\cal A} \in {\cal A} ~|~ u_{\cal A}u_{\cal A}^{\dag}=u_{\cal A}^{\dag}u_{\cal A}=1_{\cal A}\},\\
U_{\cal B}&=&\{u_{\cal B} \in {\cal B} ~|~ u_{\cal B}u_{\cal B}^{\dag}=u_{\cal B}^{\dag}u_{\cal B}=1_{\cal B}\},
\end{eqnarray}
respectively, where $1_{\cal A}$ and $1_{\cal B}$ are the units of ${\cal A}$ and ${\cal B}$. Because the set of all unitary quaternions is $SU(2)$, the gauge group acting on the fermion fields is $U(1) \times U(1)$\\$\times SU(2) \times U(3)$ (see Eqs.~(\ref{A}) and (\ref{B})). This is reduced to $U(1)_{Y}\times SU(2)_{L} \times SU(3)_{c}$ by imposing a condition later.

The connections (or covariant derivatives) $\nabla_{\cal A}$ on ${\cal E_{A}}$ and $\nabla_{\cal B}$ on ${\cal E_{B}}$ are defined by
\begin{eqnarray}
\nabla_{\cal A} \xi_{\cal A}&=&d\xi_{\cal A}+\rho_{\cal A} \xi_{\cal A} ,~~\xi_{\cal A} \in {\cal E_{A}},\\
\nabla_{\cal B} \xi_{\cal B}&=&d\xi_{\cal B}+\rho_{\cal B} \xi_{\cal B} ,~~\xi_{\cal B} \in {\cal E_{B}},
\end{eqnarray}
respectively, where $\rho_{\cal A}$ and $\rho_{\cal B}$ are the connection 1-forms on ${\cal E_{A}}$ and ${\cal E_{B}}$. The connection 1-forms are composed of elements of ${\cal A}$ and ${\cal B}$,
\begin{eqnarray}
\rho_{\cal A}&=&\sum_{s=1}^{m}a_{s}da^{\dag}_{s},~a_{s} \in {\cal A},\\
\rho_{\cal B}&=&\sum_{s=1}^{m}b_{s}db^{\dag}_{s},~b_{s} \in {\cal B},
\end{eqnarray}
where $m\ge 2$ is not specified, and
\begin{equation}
\sum_{s=1}^{m}a_{s}a^{\dag}_{s}=1_{\cal A},~~\sum_{s=1}^{m}b_{s}b^{\dag}_{s}=1_{\cal B}
\end{equation}
are required. Gauge transformations of $a_{s}$ and $b_{s}$ are defined by
\begin{eqnarray}
a_{s} &\to& u_{\cal A}a_{s},~~u_{\cal A} \in U_{\cal A},\\
b_{s} &\to& u_{\cal B}b_{s},~~u_{\cal B} \in U_{\cal B}.
\end{eqnarray}
Corresponding to the decomposition of $d$, the connection 1-forms are decomposed as
\begin{equation}
\rho_{\cal A}=\rho^{M}_{\cal A}+\rho^{F}_{\cal A},~~\rho_{\cal B}=\rho^{M}_{\cal B}+\rho^{F}_{\cal B},
\end{equation}
where
\begin{eqnarray}
\rho^{M}_{\cal A}&=&\sum_{s=1}^{m}a_{s}d^{M}a^{\dag}_{s},~~\rho^{F}_{\cal A}=\sum_{s=1}^{m}a_{s}d^{F}a^{\dag}_{s},\\
\rho^{M}_{\cal B}&=&\sum_{s=1}^{m}b_{s}d^{M}b^{\dag}_{s},~~\rho^{F}_{\cal B}=\sum_{s=1}^{m}b_{s}d^{F}b^{\dag}_{s}.
\end{eqnarray}
Representations of the components of $\rho_{\cal A}$ and $\rho_{\cal B}$ are
\begin{eqnarray}
\pi_{l}(\rho_{\cal A}^{M})&=&-i{\cal W_{A}}_{l\mu}\otimes \gamma^{\mu},~~{\cal W_{A}}_{l\mu}=\pmatrix{A_{\mu}&\begin{array}{cc}0&0\end{array}\cr \begin{array}{c}0\\0\end{array}&{\bf W}_{\mu}\cr}\otimes 1_{N},\\
\pi_{l}(\rho_{\cal A}^{F})&=&\pmatrix{0&M_{e}^{\dag}\varphi^{\dag}\cr M_{e}\varphi & \begin{array}{cc}0&0\\ 0&0 \end{array}\cr}\otimes \gamma^{5},\\
\pi_{q}(\rho_{\cal A}^{M})&=&-i{\cal W_{A}}_{q\mu}\otimes \gamma^{\mu},~~{\cal W_{A}}_{q\mu}=\pmatrix{\begin{array}{cc}A_{\mu}&0\\ 0&-A_{\mu}\end{array}&\begin{array}{cc}0&0\\0&0\end{array}\cr \begin{array}{cc}0&0\\0&0\end{array}&{\bf W}_{\mu}\cr }\nonumber \\
&&\mbox{\hspace*{4.2cm}}\otimes {\bf 1}_{3}\otimes 1_{N},\\
\pi_{q}(\rho_{\cal A}^{F})&=&\pmatrix{\begin{array}{cc}0&0\\ 0&0 \end{array}&\begin{array}{c}M_{d}^{\dag}\varphi^{\dag}\\M_{u}^{\dag}{\tilde \varphi}^{\dag}\end{array}\cr\begin{array}{cc}M_{d}\varphi &M_{u}{\tilde \varphi}\end{array}& \begin{array}{cc}0&0\\ 0&0 \end{array}\cr}\otimes {\bf 1}_{3}\otimes \gamma^{5},\\
\pi_{l}(\rho_{\cal B}^{M})&=&-i{\cal W_{B}}_{l\mu}\otimes \gamma^{\mu},~~{\cal W_{B}}_{l\mu}=B_{\mu}1_{l}\otimes 1_{N},\\
\pi_{q}(\rho_{\cal B}^{M})&=&-{\cal W_{B}}_{q\mu}\otimes \gamma^{\mu},~~{\cal W_{B}}_{q\mu}=1_{q}\otimes {\bf G}'_{\mu}\otimes 1_{N},\\
\pi_{l}(\rho_{\cal B}^{F})&=&\pi_{q}(\rho_{\cal B}^{F})=0.
\end{eqnarray}
Here, $A_{\mu},~{\bf W}_{\mu}$, $\varphi=(\varphi_{1}, \varphi_{2})^{T}$, ${\tilde \varphi}=-i\bsigma^{2}\varphi^{\ast}$, $B_{\mu}$ and ${\bf G}'_{\mu}$ are composed of matrix elements of $\pi(a_{s})$ and $\pi(b_{s})$ as follows:\footnote{$tr(\frac{\bsigma^{a}}{2}\frac{\bsigma^{b}}{2})=\frac{\delta^{ab}}{2},~~a,b=1,2,3$.}
\begin{eqnarray}
A_{\mu}&=&i\sum_{s=1}^{m}f_{s}\partial_{\mu}f^{\ast}_{s},\\
{\bf W}_{\mu}&=&i\sum_{s=1}^{m}{\bf u}_{s}\partial_{\mu}{\bf u}_{s}^{\dag},\\
\varphi_{1}&=&\sum_{s=1}^{m}\alpha_{s}f_{s}^{\ast}-1,\\
\varphi_{2}&=&-\sum_{s=1}^{m}\beta^{\ast}_{s}f_{s}^{\ast},\\
B_{\mu}&=&i\sum_{s=1}^{m}g_{s}\partial_{\mu}g^{\ast}_{s},\\
{\bf G}'_{\mu}&=&i\sum_{s=1}^{m}{\bf g}_{s}\partial_{\mu}{\bf g}_{s}^{\dag},
\end{eqnarray}
where we have defined matrix elements of $\pi(a_{s})$ and $\pi(b_{s})$ by putting the subscript $s$ on $\pi(a)$ and $\pi(b)$ of Eqs.~(\ref{component1}) and (\ref{component2}), respectively, and written
\begin{equation}
{\bf u}_{s}=\pmatrix{\alpha_{s} &\beta_{s} \cr -\beta^{\ast}_{s}&\alpha^{\ast}_{s}\cr}.
\end{equation}

To reduce the gauge group acting on the fermion fields from $U(1) \times U(1)$\\$\times SU(2) \times U(3)$ to $U(1)_{Y}\times SU(2)_{L} \times SU(3)_{c}$, the unimodularity condition
\begin{equation}
tr[E\{\pi(\rho_{\cal A})+\pi(\rho_{\cal B})\}]=0\label{unimodu}
\end{equation}
is required, where $tr$ is the ordinary trace, and
\begin{equation}
E=\pmatrix{\pmatrix{E_{R}&0&0\cr 0&E_{L}&0\cr 0&0&E_{L}\cr}\otimes 1_{N}&0\cr 0&\pmatrix{E_{R}&0&0&0\cr 0&E_{R}&0&0\cr 0&0&E_{L}&0 \cr 0&0&0&E_{L}\cr}\otimes {\bf 1}_{3}\otimes 1_{N}\cr}.\label{E}
\end{equation}
Here, the arbitrary constants $E_{R}$ and $E_{L}$ are independent of each other. This condition requires
\begin{equation}
B_{\mu}=A_{\mu}=-3G_{\mu}^{0},\label{independent}
\end{equation}
where $G_{\mu}^{0}$ is the $u(1)$ component of ${\bf G}'_{\mu}$. We define
\begin{equation}
{\bf G}'_{\mu}=G_{\mu}^{0}{\bf 1}_{3}+{\bf G}_{\mu}.
\end{equation}
Because of Eq.~(\ref{independent}), there is only a single independent $U(1)$ gauge field. We choose $B_{\mu}$ as this field.

The curvature 2-forms of $\rho_{\cal A}$ and $\rho_{\cal B}$ are defined by
\begin{equation}
\theta_{\cal A}=d\rho_{\cal A} +\rho_{\cal A}^{2},~~\theta_{\cal B}=d\rho_{\cal B} +\rho_{\cal B}^{2}.
\end{equation}
Corresponding to the decomposition of $d$, these can also be decomposed as
\begin{eqnarray}
\theta_{\cal A}&=&\theta^{MM}_{\cal A}+\theta^{MF}_{\cal A}+\theta^{FF}_{\cal A},\\
\theta_{\cal B}&=&\theta^{MM}_{\cal B}+\theta^{MF}_{\cal B}+\theta^{FF}_{\cal B},
\end{eqnarray}
where
\begin{eqnarray}
\theta^{MM}_{\cal A,B}&=&d^{M}\rho^{M}_{\cal A,B}+\rho^{M2}_{\cal A,B},\\
\theta^{MF}_{\cal A,B}&=&d^{M}\rho^{F}_{\cal A,B}+\rho^{M}_{\cal A,B}\rho^{F}_{\cal A,B}+d^{F}\rho^{M}_{\cal A,B}+\rho^{F}_{\cal A,B}\rho^{M}_{\cal A,B},\\
\theta^{FF}_{\cal A,B}&=&d^{F}\rho^{F}_{\cal A,B}+\rho^{F2}_{\cal A,B}.
\end{eqnarray}
Representations of the components of $\theta_{\cal A}$ and $\theta_{\cal B}$ are
\begin{eqnarray}
\pi_{l,q}(\theta^{MM}_{\cal A,B})&=&-\frac{1}{2}(\partial^{[\mu}{\cal W}_{{\cal A,B}l,q}^{\nu ]}-i[{\cal W}_{{\cal A,B}l,q}^{\mu},{\cal W}_{{\cal A,B}l,q}^{\nu}])\otimes \sigma_{\mu \nu}+Y_{{\cal A,B}l,q}\otimes 1_{C},\\
\pi_{l}(\theta^{MF}_{\cal A})&=&\pmatrix{0&(M_{e}{\bf D}^{\phi}_{\mu}\phi)^{\dag}\cr M_{e}{\bf D}^{\phi}_{\mu}\phi&\begin{array}{cc}0&0\\ 0&0 \end{array}\cr}\otimes \gamma^{\mu}\gamma^{5},\\
\pi_{q}(\theta^{MF}_{\cal A})&=&\pmatrix{0&0&(M_{d}{\bf D}^{\phi}_{\mu}\phi)^{\dag}\cr 0&0&(M_{u}{\bf D}^{\tilde \phi}_{\mu}{\tilde \phi})^{\dag}\cr M_{d}{\bf D}^{\phi}_{\mu}\phi&M_{u}{\bf D}^{\tilde \phi}_{\mu}{\tilde \phi}&\begin{array}{cc}0&0\\ 0&0 \end{array}\cr}\otimes {\bf 1}_{3}\otimes \gamma^{\mu}\gamma^{5},\\
\pi_{l,q}(\theta^{MF}_{\cal B})&=&0,\\
\pi_{l}(\theta^{FF}_{\cal A})&=&\pmatrix{M_{e}^{\dag}M_{e}(\phi^{\dag}\phi-1)&\begin{array}{cc}0&0\end{array}\cr \begin{array}{c}0\\0\end{array}&M_{e}M_{e}^{\dag}\left(\phi\phi^{\dag}-\pmatrix{1&0\cr 0&0\cr}+{\bf Z}\right)\cr}\otimes 1_{C},\\
\pi_{q}(\theta^{FF}_{\cal A})&=&\pmatrix{A&\begin{array}{cc}0&0\\ 0&0\end{array}\cr \begin{array}{cc}0&0\\ 0&0\end{array}&B\cr}\otimes {\bf 1}_{3}\otimes 1_{C},\\
A&=&\pmatrix{M_{d}^{\dag}M_{d}(\phi^{\dag}\phi-1)&0\cr 0&M_{u}^{\dag}M_{u}(\phi^{\dag}\phi-1)\cr},\nonumber \\
B&=&M_{d}M_{d}^{\dag}\left(\phi \phi^{\dag}-\pmatrix{1&0\cr 0&0\cr}+{\bf Z}\right)+M_{u}M_{u}^{\dag}\left({\tilde \phi} {\tilde \phi}^{\dag}-\pmatrix{0&0\cr 0&1\cr}-{\bf Z}\right),\nonumber \\
\pi_{l,q}(\theta^{FF}_{\cal B})&=&0,
\end{eqnarray}
where
\begin{eqnarray}
\sigma_{\mu \nu}&=&\frac{i}{2}[\gamma_{\mu},\gamma_{\nu}],\\
Y_{{\cal A,B}l,q}&=&-i(\partial_{\mu}{\cal W}_{{\cal A,B}l,q}^{\mu}-i{\cal W}_{{\cal A,B}l,q\mu}{\cal W}_{{\cal A,B}l,q}^{\mu}+X_{{\cal A,B}l,q}),\\
X_{{\cal A}l,q}&=&-i\sum_{s=1}^{m}\pi_{l,q}(a_{s})\partial^{2}\pi_{l,q}(a_{s}^{\dag}),\\
X_{{\cal B}l,q}&=&-i\sum_{s=1}^{m}\pi_{l,q}(b_{s})\partial^{2}\pi_{l,q}(b_{s}^{\dag}),\\
\phi&=&\varphi +\pmatrix{1\cr 0\cr },~{\tilde \phi}=-i\bsigma^{2}\phi^{\ast},\\
D^{\phi}_{\mu}\phi&=&\partial_{\mu}\phi-i({\bf W}_{\mu}\phi-\phi A_{\mu}),\\
D^{\tilde \phi}_{\mu}{\tilde \phi}&=&\partial_{\mu}{\tilde \phi}-i({\bf W}_{\mu}{\tilde \phi}+{\tilde \phi}A_{\mu}),\\
{\bf Z}&=&\sum_{s=1}^{m}\pmatrix{\beta_{s}\beta_{s}^{\ast}&\alpha_{s}\beta_{s}\cr \alpha_{s}^{\ast}\beta_{s}^{\ast}&-\beta_{s}\beta_{s}^{\ast}\cr}.
\end{eqnarray}

The lagrangian density of the fermion fields is defined by
\begin{equation}
{\cal L}_{f}={\bar \Psi}\{D+\pi(\rho_{\cal A})+\pi(\rho_{\cal B})\}\Psi,\label{L_f}
\end{equation}
where ${\bar \Psi}=\Psi^{\dag}\gamma^{4}=\Psi^{\dag}i\gamma^{0}$.
Through the re-definitions
\begin{eqnarray}
B_{\mu}&\to &-\frac{1}{2}g'B_{\mu},\label{redefinition1}\\
{\bf W}_{\mu}&\to &g_{2}\pmatrix{0&1\cr 1&0\cr}{\bf W}_{\mu}\pmatrix{0&1\cr 1&0\cr},\\
{\bf G}_{\mu}&\to &g_{3}{\bf G}_{\mu},\\
\phi&\to &\frac{\sqrt{2}}{v}\pmatrix{0&1\cr 1&0\cr}\phi,~~~(\tilde{\phi}=-i\bsigma_{2}\phi^{\ast}\to \tilde{\phi}=i\bsigma_{2}\phi^{\ast}),\label{redefinition4}\\
\Psi&\to &\exp{\left(-i\frac{3\pi}{4}\gamma^{5}\right)}\Psi,
\end{eqnarray}
we obtain the fermionic lagrangian density of the standard model. Here, $g'~g_{2}$ and $g_{3}$ are gauge coupling constants of $U(1)_{Y},~SU(2)_{L}$ and $SU(3)_{c}$, respectively, and $B_{\mu},~{\bf W}_{\mu}=\sum_{a=1 \sim 3}(\bsigma^{a}/2)W_{a \mu}$ and ${\bf G}_{\mu}=\sum_{a=1 \sim 8}(\blambda^{a}/2)G_{a \mu}$\footnote{$tr(\frac{\blambda^{a}}{2}\frac{\blambda^{b}}{2})=\frac{\delta^{ab}}{2},~~a,b=1 \sim 8$.} \ are respective gauge fields, and $\phi$ is the Higgs doublet. The matrices $h_{a}~~(a=e,u,d)$ of the Yukawa coupling constants are expressed as
\begin{equation}
h_{a}=\frac{\sqrt{2}}{v}M_{a},\label{Yukawa coupling}
\end{equation}
where $v$ is an arbitrary parameter that can be introduced. We will see later that $v$ corresponds to the vacuum expectation value of the Higgs field.

The original form of the bosonic lagrangian density defined by Connes and Lott is\cite{1,2}
\begin{equation}
{\cal L}_{\mbox{{\scriptsize Connes}}}=tr\{K_{\cal A}\pi(\theta_{\cal A}^{2})+K_{\cal B}\pi(\theta_{\cal B}^{2})\},\label{Connes}
\end{equation}
where $K_{\cal A}$ and $K_{\cal B}$ are any matrices that commute with all elements of the gauge groups $\pi(U_{\cal A})$ and $\pi(U_{\cal B})$, respectively. This lagrangian density contains extra fields $X_{{\cal A,B}l,q}$ and ${\bf Z}$, which have no kinetic terms. After elimination of the fields by using their equations of motion, and the redefinitions (\ref{redefinition1}) $\sim$ (\ref{redefinition4}), we obtain the bosonic lagrangian density of the standard model. Then we also need to adjust the matrix elements of $K_{\cal A,B}$ to give the correct coefficients to the kinetic terms of the gauge and Higgs fields. Constraints among coupling constants of the model are derived from specific restrictions on the matrix forms of $K_{\cal A,B}$. Connes considered\cite{2}
\begin{equation}
K_{\cal A} \in \pi({\cal A}),~~K_{\cal B} \in \pi({\cal B}),
\end{equation}
and then a constraint among coupling constants is required.

Now, we note that $\pi(\theta^{MM}_{\cal A,B}),~\pi(\theta^{MF}_{\cal A})$ and $\pi(\theta^{FF}_{\cal A})$ do not mix with each other under the gauge and Lorentz transformations. Thus, we can generalize Eq.~(\ref{Connes}) as\footnote{$\pi (\theta^{MF2}_{\cal B})=\pi (\theta^{FF2}_{\cal B})=0$.}
\begin{equation}
{\cal L}'=-\frac{1}{g_{2}^{2}}tr\{K_{{\cal A}1}\pi(\theta^{MM2}_{\cal A})+K_{{\cal A}2}\pi(\theta^{MF2}_{\cal A})+K_{{\cal A}3}\pi(\theta^{FF2}_{\cal A})+K_{\cal B}\pi(\theta^{MM2}_{\cal B})\},\label{Lb}
\end{equation}
where $K_{{\cal A}i=1,2,3}$ and $K_{\cal B}$ are matrices that commute with all elements of the gauge groups $\pi(U_{\cal A})$ and $\pi(U_{\cal B})$, respectively. We take
\begin{equation}
K_{{\cal A}i=1,2,3} \in \pi({\cal A}),~~K_{\cal B} \in \pi({\cal B})\label{KA KB}
\end{equation}
to be the same as Connes's construction and write
\begin{eqnarray}
\pi(K_{{\cal A}i})&=&\pmatrix{\pmatrix{\kappa_{iR}&0&0\cr 0&\kappa_{iL}&0\cr 0&0&\kappa_{iL}\cr}&0\cr 0&\pmatrix{\kappa_{iR}&0&0&0\cr 0&\kappa_{iR}&0&0\cr 0&0&\kappa_{iL}&0 \cr 0&0&0&\kappa_{iL}\cr}\otimes {\bf 1}_{3}\cr}\otimes 1_{N},~~~~~~~~\label{KAi}\\
\pi(K_{\cal B})&=&\pmatrix{\pmatrix{\kappa_{l}&0&0\cr 0&\kappa_{l}&0\cr 0&0&\kappa_{l}\cr}&0\cr 0&\pmatrix{\kappa_{q}&0&0&0\cr 0&\kappa_{q}&0&0\cr 0&0&\kappa_{q}&0 \cr 0&0&0&\kappa_{q}\cr}\otimes {\bf 1}_{3}\cr}\otimes 1_{N}.
\end{eqnarray}
We will see that these forms are sufficient to give the most general lagrangian density of the boson fields. After elimination of the extra fields $X_{{\cal A,B}l,q}$ and ${\bf Z}$ from ${\cal L}'$, and the redefinitions (\ref{redefinition1}) $\sim$ (\ref{redefinition4}), we obtain
\begin{eqnarray}
{\cal L}&=&-C_{1}F_{\mu \nu}^{B}F^{B\mu \nu}-C_{2}F_{\mu \nu a}^{W}F^{W \mu \nu a}-C_{3}G_{\mu \nu a}G^{\mu \nu a}\nonumber \\
&&+C_{4}(D_{\mu}\phi)^{\dag}D^{\mu}\phi-C_{5}\left(\phi^{\dag}\phi-\frac{v^{2}}{2}\right)^{2},
\end{eqnarray}
where
\begin{eqnarray}
C_{1}&=&N\left(\frac{g'}{g_{2}}\right)^{2}\left(\frac{7}{2}\kappa_{1R}+\frac{3}{2}\kappa_{l}+\frac{2}{3}\kappa_{q}\right),\\
C_{2}&=&4N\kappa_{1L},\\
C_{3}&=&4N\kappa_{q}\left(\frac{g_{3}}{g_{2}}\right)^{2},\\
C_{4}&=&\frac{8}{g_{2}^{2}v^{2}}(\kappa_{2R}+\kappa_{2L})tr(M_{e}^{\dag}M_{e}+3M_{u}^{\dag}M_{u}+3M_{d}^{\dag}M_{d}),~~~~~\\
C_{5}&=&\frac{16}{v^{4}g_{2}^{2}}[\kappa_{3R}tr\{(M_{e}M_{e}^{\dag})^{2}+3(M_{u}M_{u}^{\dag})^{2}+3(M_{d}M_{d}^{\dag})^{2}\}\nonumber \\
&&+\frac{1}{2}\kappa_{3L}tr\{(M_{e}M_{e}^{\dag})^{2}+3(M_{d}M_{d}^{\dag}+M_{u}M_{u}^{\dag})^{2}\}].
\end{eqnarray}
Here, $F_{\mu \nu}^{B}$, $F_{\mu \nu a}^{W}$ and $G_{\mu \nu a}$ are the field strengths of $B_{\mu}$, $W_{\mu a}$ and $G_{\mu a}$, respectively, and $D^{\mu}\phi=\partial_{\mu}\phi-i(g_{2}{\bf W}_{\mu}\phi+\frac{g'}{2}B_{\mu}\phi )$. It is necessary that
\begin{eqnarray}
C_{1}&=&C_{2}=C_{3}=\frac{1}{4},\label{requirement-1}\\
C_{4}&=&1,\label{requirement-4}\\
C_{5}&=&\lambda,\label{requirement-5}
\end{eqnarray}
where $\lambda$ is the quartic self-coupling constant of the Higgs field. Thus we have
\begin{eqnarray}
\alpha_{1}&=&\frac{5}{2N(21\kappa_{1R}+9\kappa_{l}+4\kappa_{q})}\alpha_{2},\label{f1}\\
1&=&\frac{1}{16N\kappa_{1L}},\label{f2}\\
\alpha_{3}&=&\frac{1}{16N\kappa_{q}}\alpha_{2},\label{f3}\\
\alpha_{T}&=&\frac{1}{12(\kappa_{2R}+\kappa_{2L})}\alpha_{2},\label{fT}\\
\alpha_{\lambda}&=&\frac{2\kappa_{3R}+\kappa_{3L}}{24(\kappa_{2R}+\kappa_{2L})}\alpha_{2}=6(2\kappa_{3R}+\kappa_{3L})\frac{\alpha_{T}^{2}}{\alpha_{2}},\label{flambda}
\end{eqnarray}
where
\begin{equation}
\alpha_{i}=\frac{g_{i}^{2}}{4\pi}~\left(i=1,2,3,~~g_{1}=\sqrt{\frac{5}{3}}g'\right),~~\alpha_{T}=\frac{h_{T}^{2}}{4\pi},~~\alpha_{\lambda}=\frac{\lambda}{4\pi},
\end{equation}
and $h_{T}$ is the Yukawa coupling constant of the $N$-th generation up quark. To obtain these relations we have also used Eq.~(\ref{Yukawa coupling}), and assumed that non-diagonal parts of $h_{a}$ of Eq.~(\ref{Yukawa coupling}) and Yukawa coupling constants other than $h_{T}$ are negligible. We note that we can give any values to $\alpha_{i=1,2,3,T,\lambda}$ under Eqs.~(\ref{f1}) $\sim$ (\ref{flambda}), because $\kappa_{iR,iL}$ and $\kappa_{l,q}$ are independent of each other. From this result, we realize that Eqs.~(\ref{f1}) $\sim$ (\ref{flambda}) are not constraints among coupling constants, but give a new parametrization of the coupling constants of the standard model. Constraints are obtained by imposing some restrictions on the matrix forms of $K_{{\cal A}i=1,2,3}$ and $K_{\cal B}$. We will consider two kinds of constraints in \S \S 3 and 4.
\section{A renormalization group analysis of the Higgs boson mass}

In this section we give certain constraints among coupling constants by imposing  natural restrictions on $K_{{\cal A}i=1,2,3}$ and $K_{\cal B}$. We adopt the idea of \'Alvarez et al., and analyze the renormalization group evolution of the Higgs boson mass by using the constraints.

First we put
\begin{equation}
K_{{\cal A}1}=K_{{\cal A}2}=K_{{\cal A}3}\equiv K_{\cal A}.\label{restriction on K's}
\end{equation}
Then we have a similar model that of Connes.\cite{2} \ In this case we have four independent parameters, $\kappa_{R}, \kappa_{L}, \kappa_{l}$ and $ \kappa_{q}$, where we have defined the components of $K_{\cal A}$ by removing the subscript $i$ from components of $K_{{\cal A}i}$ of Eq.~(\ref{KAi}). Then Eqs.~(\ref{f1}) $\sim$ (\ref{flambda}) give rise to a certain constraint, by which $\alpha_{\lambda}$ can be expressed in terms of other coupling constants. This circumstance is the same as that in Connes's model. To determine the RG evolution of the Higgs boson mass completely, we need one more constraint that is used to determine the energy scale at which the NCG constraints hold. So we consider further restriction on $K_{\cal A,B}$.

We consider a restriction on the algebras $\cal A$ and $\cal B$ defined by Eqs.~(\ref{A}) and (\ref{B}). We write the elements of ${\cal A}$ as
\begin{equation}
(f, {\bf u}),
\end{equation}
where 
\begin{equation}
f \in C^{\infty}(M)\otimes {\bf C},~~{\bf u} \in C^{\infty}(M) \otimes {\bf H}.
\end{equation}
We also write the elements of ${\cal B}$ as
\begin{equation}
(g, {\bf g}),
\end{equation}
where 
\begin{equation}
g \in C^{\infty}(M) \otimes {\bf C},~~{\bf g} \in C^{\infty}(M)\otimes M_{3}({\bf C}).
\end{equation}
In Connes's construction, $f$ and $g$ can be taken independently. Here we suppose that we must always specify the elements of $\cal A$ and $\cal B$ at the same time, and take
\begin{equation}
f=g.
\end{equation}
In this case,
\begin{equation}
B_{\mu}=A_{\mu}
\end{equation}
of Eq.~(\ref{independent}) holds automatically without the unimodularity condition (\ref{unimodu}), and also
\begin{equation}
B_{\mu}=-3G_{\mu}
\end{equation}
of Eq.~(\ref{independent}) can be obtained from the condition
\begin{equation}
tr\{\pi(\rho_{\cal A})+\pi(\rho_{\cal B})\}=0,
\end{equation}
which is the usual unimodularity condition.

As a result of the restriction on the algebras $\cal A$ and $\cal B$, we have restrictions on the forms of $K_{{\cal A}} \in {\cal A},~~ K_{\cal B}\in {\cal B}$:
\begin{equation}
\kappa_{R}=\kappa_{l} \equiv \kappa_{c}.\label{kc}
\end{equation}
In this case, the independent parameters are $\kappa_{c}, \kappa_{L}$ and $\kappa_{q}$. Then Eqs.~(\ref{f1}) $\sim$ (\ref{flambda}) give rise to two constraints,
\begin{eqnarray}
\alpha_{t}&=&\left(\frac{1}{3\alpha_{1}}+\frac{1}{4\alpha_{2}}-\frac{1}{30\alpha_{3}}\right)^{-1},\label{first constraint}\\
\alpha_{\lambda}&=&\left(\frac{1}{3\alpha_{1}}+\frac{1}{8\alpha_{2}}-\frac{1}{30\alpha_{3}}\right)\alpha_{t}^{2},\label{second constraint}
\end{eqnarray}
where we have put $N=3$, and $\alpha_{t}=h_{t}^{2}/4\pi$ is the Yukawa coupling constant of the top quark.

Let us study the RG evolution of the Higgs boson mass by using the constraints (\ref{first constraint}) and (\ref{second constraint}). According to \'Alvarez et al., we consider that the constraints hold only at a certain energy scale $\mu_{0}$. We determine the energy scale by using Eq.~(\ref{first constraint}) as follows. The RG evolution of $\alpha_{i=1,2,3,t}$ can be known by solving
\begin{equation}
\frac{d\alpha_{i}}{dt}=\beta_{\alpha_{i}},~~(i=1,2,3,t)\label{def beta}
\end{equation}
where $t=\ln(\mu/m_{Z})~(m_{Z}=91.2 \mbox{~GeV})$, and the $\beta$ functions are
\begin{eqnarray}
4\pi\beta_{\alpha_{1}}&=&\left(\frac{8N}{3}+\frac{1}{5}\right)\alpha_{1}^{2},\label{1-loop b of a1}\\
4\pi\beta_{\alpha_{2}}&=&\left(\frac{8N}{3}-\frac{43}{3}\right)\alpha_{2}^{2},\\
4\pi\beta_{\alpha_{3}}&=&\left(\frac{8N}{3}-22\right)\alpha_{3}^{2},\label{1-loop b of a3}\\
4\pi \beta_{\alpha_{t}}&=&\alpha_{t}\left(9\alpha_{t}-\frac{17}{10}\alpha_{1}-\frac{9}{2}\alpha_{2}-16\alpha_{3}\right)\label{1-loop b of at}
\end{eqnarray}
in the 1-loop approximation.\cite{15,16} \ (We put $N=3$.) For $i=1,2,3$, the solutions are\cite{v,12,13}
\begin{eqnarray}
\alpha_{1}^{-1}(t)&=&\alpha_{1}^{-1}(t=0)-\frac{41}{20\pi}t,\label{solution-1}\\
\alpha_{2}^{-1}(t)&=&\alpha_{2}^{-1}(t=0)+\frac{19}{12\pi}t,\\
\alpha_{3}^{-1}(t)&=&\alpha_{3}^{-1}(t=0)+\frac{7}{2\pi}t,\label{solution-2}
\end{eqnarray}
where
\begin{equation}
\alpha_{1}(t=0)=0.01698,~~\alpha_{2}(t=0)=0.03364,~~\alpha_{3}(t=0)=0.12.\label{a123}
\end{equation}
We can also solve Eq.~(\ref{def beta}) for $i=t$ with the initial condition\cite{12}
\begin{equation}
\alpha_{t}(t=0)=0.09283.
\end{equation}
The evolution of $\alpha_{t}$ and the right-hand side of Eq.~(\ref{first constraint}) are written in Fig.~1.
\begin{figure}
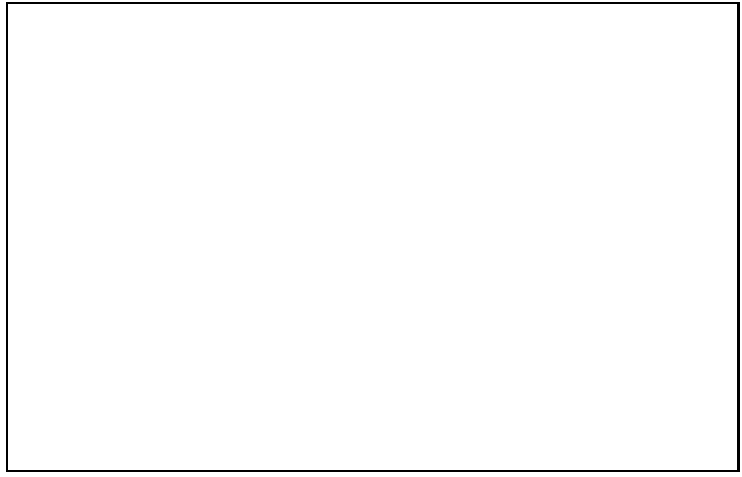

\begin{center}
\fbox{\begin{minipage}{95mm} \vspace*{60mm} \hspace*{92mm} \end{minipage}}
\end{center}
\caption{The RG evolution of $\alpha_{t}$ and the right-hand side of Eq.~(3$\cdot$11). Equation (3$\cdot$11) holds at $t_{0}=15.26$.}
\label{fig:1}
\end{figure}
From the figure we can determine the energy scale at which Eq.~(\ref{first constraint}) holds:
\begin{equation}
\mu_{0}=3.87 \times 10^{8} \mbox{GeV}.~~(t_{0}=\ln(\mu_{0}/m_{Z})=15.26)
\end{equation}

Equation (\ref{second constraint}) also holds at the same energy scale. So, by estimating the right-hand side of Eq.~(\ref{second constraint}) at this scale, we obtain
\begin{equation}
\alpha_{\lambda}(t_{0}=15.26)=0.03269.
\end{equation}
Using this as the initial condition, we can solve
\begin{equation}
\frac{d\alpha_{\lambda}}{dt}=\beta_{\alpha_{\lambda}}
\end{equation}
numerically, where the $\beta$ function is
\begin{equation}
4\pi \beta_{\alpha_{\lambda}}=24\alpha_{\lambda}^{2}+ 12\alpha_{\lambda}\alpha_{t}-\frac{9}{5}\alpha_{\lambda}\alpha_{1}-9\alpha_{\lambda}\alpha_{2}+\frac{27}{200}\alpha_{1}^{2}+\frac{9}{20}\alpha_{1}\alpha_{2}+\frac{9}{8}\alpha_{2}^{2}-6\alpha_{t}^{2}
\end{equation}
in the 1-loop approximation.\cite{15,16} \ The Higgs boson mass is expressed as
\begin{equation}
m_H=\sqrt{8\pi v^{2}\alpha_{\lambda}}.
\end{equation}
It is known that
\begin{equation}
v(\mu=m_{Z})=246 \mbox{~GeV},
\end{equation}
and its running effect is negligible.\cite{v,12,13} \ Thus, we can determine the RG evolution of the Higgs boson mass. It appears in Fig.~2.
\begin{figure}
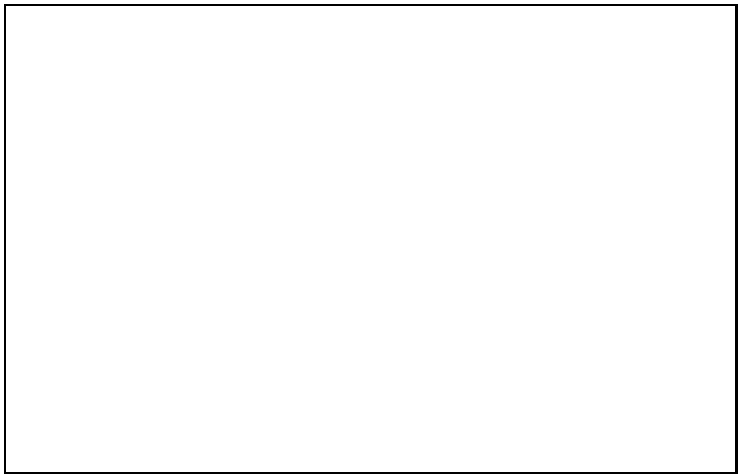

\begin{center}
\fbox{\begin{minipage}{95mm} \vspace*{60mm} \hspace*{92mm} \end{minipage}}
\end{center}
\caption{The RG evolution of $m_H(t)$ and the curve of $\mu=m_{Z}e^{t}$. $m_H(\mu=m_H)$ can be read from the intersecting point of them.}
\label{fig:2}
\end{figure}
The figure shows $m_H(\mu=m_H)=195 \mbox{~GeV}$.
\section{RGI constraints in the NCG method}
In this section, we consider RGI constraints in the NCG parametrization (\ref{f1}) $\sim$ (\ref{flambda}). First we consider possible RGI constraints among coupling constants of the standard model. We consider the form
\begin{equation}
\alpha_{i}=F_{i}(\alpha_{2})\alpha_{2},~~(i=1,3,T,\lambda)\label{fi}
\end{equation}
where $F_{i}(\alpha_{2})$ are certain functions of $\alpha_{2}$. The conditions of the constraints (\ref{fi}) being RGI are obtained by differentiating bothsides of Eq.~(\ref{fi}) by the scale parameter $\mu$. We obtain
\begin{equation}
\frac{dF_{i}(s)}{ds}=-F_{i}(s)+\frac{\beta_{\alpha_{i}}}{\beta_{\alpha_{2}}},~~(i=1,3,T,\lambda)\label{RGI constraint}
\end{equation}
where
\begin{equation}
s=\ln \alpha_{2}
\end{equation}
and
\begin{equation}
\beta_{\alpha_{i}}=\mu\frac{d\alpha_{i}}{d\mu}.
\end{equation}
We can express $\beta_{\alpha_{i}}/\beta_{\alpha_{2}}$ of Eq.~(\ref{RGI constraint}) in terms of $F_{i=1,3,T,\lambda}$ by using the expressions of 1-loop $\beta$ functions with Eq.~(\ref{fi}).  In particular, we only consider the cases of $i=1,3$. For these cases, we obtain simple expressions,
\begin{equation}
\frac{\beta_{\alpha_{i}}}{\beta_{\alpha_{2}}}=b_{i}F_{i}^{2},~~(i=1,3)\label{partial dif}
\end{equation}
where
\begin{eqnarray}
b_{1}&=&\frac{8N+\frac{3}{5}}{8N-43},\\
b_{3}&=&\frac{8N+66}{8N-43}.
\end{eqnarray}
Here, we have used the 1-loop $\beta$ functions of Eqs.~(\ref{1-loop b of a1}) $\sim$ (\ref{1-loop b of a3}). The solutions of Eq.~(\ref{partial dif}) are
\begin{equation}
F_{i}(\alpha_{2})=\frac{1}{a_{i}\alpha_{2}+b_{i}},\label{solution}
\end{equation}
where $a_{i}$ are certain constants. The constants $a_{i}$ are determined such that the constraints (\ref{fi}) hold for $\alpha_{i=1,2,3}(\mu=m_{Z})$ of Eq.~(\ref{a123}). We obtain
\begin{equation}
a_{i}=\frac{1}{\alpha_{i}(\mu=m_{Z})}-\frac{b_{i}}{\alpha_{2}(\mu=m_{Z})}.
\end{equation}

It is easy to give these RGI constraints in the NCG parametrization. By comparing Eq.~(\ref{solution}) with Eqs.~(\ref{f1}) and (\ref{f3}), we obtain
\begin{eqnarray}
\kappa_{c}&=&\frac{1}{12N}\left\{\left(a_{1}-\frac{a_{3}}{10}\right)\alpha_{2}+b_{1}-\frac{b_{3}}{10}\right\},\\
\kappa_{q}&=&\frac{1}{16N}(a_{3}\alpha_{2}+b_{3}),
\end{eqnarray}
where we have taken $\kappa_{1R}=\kappa_{l} \equiv \kappa_{c}$ as in Eq.~(\ref{kc}).
\section{RGI constrains and the multiplicative renormalizability}
In this section, we consider the relation between the condition that a constraint among coupling constants of a model becomes RGI and the condition that the model becomes multiplicative renormalizable. We consider a Yukawa $+~\phi^{4}$ model in a mass-independent renormalization scheme. We use the minimal subtraction scheme with dimensional regularization.\cite{17,18,16} \  Our argument is dependent on the renormalization scheme.

First, we consider the condition of a RGI constraint. We denote the coupling constants of the $\phi^{4}$ and the Yukawa interactions as $\lambda$ and $h$, respectively. Suppose that there is a constraint
\begin{equation}
\alpha_{\lambda}=F(\alpha_{h})\alpha_{h}\label{constraint},
\end{equation}
where
\begin{equation}
\alpha_{\lambda}=\frac{\lambda}{4\pi},~~\alpha_{h}=\frac{h^{2}}{4\pi},\label{def of alpha}
\end{equation}
and $F(\alpha_{h})$ is a function of $\alpha_{h}$. The condition for the constraint to be RGI is
\begin{equation}
\frac{d}{d\alpha_{h}}\{F(\alpha_{h})\alpha_{h}\}=\frac{\beta_{\alpha_{\lambda}}}{\beta_{\alpha_{h}}},\label{RGI}
\end{equation}
where
\begin{equation}
\beta_{\alpha_{\lambda}}=\mu\frac{d\alpha_{\lambda}}{d\mu},~~\beta_{\alpha_{h}}=\mu\frac{d\alpha_{h}}{d\mu}.
\end{equation}

Next, we consider the condition required for the multiplicative renormalizability of the model whose bare coupling constants are constrained by
\begin{equation}
\alpha_{\lambda_{0}}=F(\alpha_{h_{0}})\alpha_{h_{0}}.\label{bare constraint}
\end{equation}
Here, $\alpha_{\lambda_{0}}$ and $\alpha_{h_{0}}$ are expressed in terms of the bare coupling constants $\lambda_{0}$ and $h_{0}$ of the $\phi^{4}$ and the Yukawa interactions, respectively, just as Eq.~(\ref{def of alpha}). In the dimensional regularization, we have
\begin{equation}
\lambda_{0}=\mu^{2\epsilon}Z_{\lambda}\lambda,~~h_{0}=\mu^{\epsilon}Z_{h}h,\label{0}
\end{equation}
where $\epsilon=(4-d)/2 \to 0$, and $d$ is the dimension of the spacetime. Now, let us suppose
\begin{equation}
F(\alpha_{h_{0}})=\sum_{n=0}^{\infty}g_{n}(\mu^{-2\epsilon}\alpha_{h_{0}})^{n},
\end{equation}
where $g_{n}$ are expansion coefficients. By substituting $\lambda_{0}$ and $h_{0}$ of Eq.~(\ref{0}) into Eq.~(\ref{bare constraint}), using the bare coupling constant version of Eq.~{(\ref{def of alpha}), we obtain
\begin{equation}
Z_{\lambda}\alpha_{\lambda}=\sum_{n=0}^{\infty}g_{n}Z_{h}^{2(n+1)}\alpha_{h}^{n+1}.\label{condition MR}
\end{equation}
Because each $Z_{\lambda}$ and $Z_{h}$ is a certain function of $\lambda$ and $h$ determined so as to provide the counterterm, this is the condition for $g_{n}$, i.e., the condition for the functional form of $F(\alpha_{h})$. We study the relation between this condition and the RGI condition (\ref{RGI}). Substituting the $\hbar$ expansions
\begin{equation}
Z_{\lambda,h}=1+\hbar Z_{\lambda,h}^{(1)}+\hbar^{2} Z_{\lambda,h}^{(2)}+\cdots
\end{equation}
into Eq.~(\ref{condition MR}), and collecting the coefficients in front of $\hbar^{0}$ and $\hbar^{1}$, respectively, we obtain
\begin{eqnarray}
\alpha_{\lambda}&=&F(\alpha_{h})\alpha_{h},\label{Renorm-0}\\
\frac{d}{d\alpha_{h}}\{F(\alpha_{h})\alpha_{h}\}&=&\frac{\alpha_{\lambda}Z_{\lambda}^{(1)}}{2\alpha_{h}Z_{h}^{(1)}}.\label{Renorm-1}
\end{eqnarray}
These are necessary conditions for Eq.~(\ref{condition MR}). We note that Eq.~(\ref{Renorm-0}) is the same as Eq.~(\ref{constraint}). On the other hand, Eq.~(\ref{Renorm-1}) can be reduced as follows: We can write
\begin{equation}
Z_{\lambda,h}=1+\sum_{i=1}^{\infty}\frac{a_{\lambda,h i}}{\epsilon^{i}},\label{expand Z}
\end{equation}
where $a_{\lambda,h i}$ are certain functions of $\lambda$ and $h$. We consider the $\hbar$ expansions
\begin{equation}
a_{\lambda,hi}=1+\hbar a_{\lambda,hi}^{(1)}+\hbar^{2} a_{\lambda,h i}^{(2)}+\cdots.
\end{equation}
Then we have
\begin{equation}
Z_{\lambda,h}^{(1)}=\frac{a_{\lambda,h 1}^{(1)}}{\epsilon}.\label{Z}
\end{equation}
On the other hand, we have
\begin{eqnarray}
\mu\frac{d\lambda}{d\mu}&=&\beta_{\lambda}=\lambda\left(2\lambda\frac{\partial}{\partial \lambda}+h\frac{\partial}{\partial h}\right)a_{\lambda 1},\label{beta-lambda-1}\\
\mu\frac{dh}{d\mu}&=&\beta_{h}=h\left(2\lambda\frac{\partial}{\partial \lambda}+h\frac{\partial}{\partial h}\right)a_{h1}\label{beta-h-1}
\end{eqnarray}
as a consequence of the fact that the $\beta$ functions are finite for $\epsilon \to 0$. We can also obtain
\begin{eqnarray}
a_{\lambda,h 1}^{(1)}&=&\sum_{\frac{l}{2}+m=1}a_{\lambda,h 1}^{(1)}(l,m)h^{l}\lambda^{m}\label{a(1)}\\
&=&a_{\lambda,h 1}^{(1)}(2,0)h^{2}+a_{\lambda,h 1}^{(1)}(0,1)\lambda +a_{\lambda,h 1}^{(1)}(4,-1)\frac{h^{4}}{\lambda}\nonumber
\end{eqnarray}
from 1-loop calculations, where $a_{\lambda,h 1}^{(1)}(l,m)$ are certain constants.\footnote{$a_{h 1}^{(1)}(0,1)=a_{h 1}^{(1)}(4,-1)=0$.} \ By substituting $a_{\lambda 1}^{(1)}$ and $a_{h 1}^{(1)}$ of Eq.~(\ref{a(1)}) into Eqs.~(\ref{beta-lambda-1}) and (\ref{beta-h-1}), respectively, we obtain the following expressions for the 1-loop $\beta$ functions:
\begin{eqnarray}
\beta_{\lambda}^{(1)}&=&2\lambda a_{\lambda 1}^{(1)},\label{beta-lambda-2}\\
\beta_{h}^{(1)}&=&2ha_{h 1}^{(1)}.\label{beta-h-2}
\end{eqnarray}
Due to Eqs.~(\ref{Z}), (\ref{beta-lambda-2}), (\ref{beta-h-2}) and
\begin{equation}
\beta_{\alpha_{\lambda}}=\frac{\beta_{\lambda}}{4\pi},~\beta_{\alpha_{h}}=\frac{h\beta_{h}}{2\pi},
\end{equation}
Eq.~(\ref{Renorm-1}) is reduced to
\begin{equation}
\frac{d}{d\alpha_{h}}\{F(\alpha_{h})\alpha_{h}\}=\frac{\beta_{\alpha_{\lambda}}^{(1)}}{\beta_{\alpha_{h}}^{(1)}}.
\end{equation}
This is the same as the RGI condition (\ref{RGI}) under the 1-loop approximation.

Thus, we can conclude that the condition that the constraint (\ref{constraint}) becomes RGI is a necessary condition of the multiplicative renormalizability of the model whose bare coupling constants are constrained by Eq.~(\ref{bare constraint}).

\section{Conclusion and discussion}
We have studied the constraints among coupling constants of the standard model obtained in the NCG method. We knew that we could obtain even the most general lagrangian of the model by generalizing the manner of constructing the lagrangian from the field strengths defined in NCG. In this sense, we can say that the NCG method gives a parametrization of the coupling constants of the standard model different from the one usually used. NCG constraints are obtained by imposing some restrictions on the way of constructing the bosonic lagrangian. Based on this observation, we made the following two considerations.

First, we formulated two constraints by imposing certain restrictions on the matrix forms of $K_{{\cal A}i=1,2,3} \in \pi({\cal A})$ and $K_{\cal B} \in \pi({\cal B})$ that are parameter matrices contained in the bosonic lagrangian. The restrictions can be regarded as a result of certain restriction on the algebras ${\cal A}$ and ${\cal B}$. When we construct the standard model with the restricted algebras, the procedure of reducing the three $U(1)$ gauge fields into the $U(1)_{Y}$ gauge field, which has been done by imposing a unimodularity condition in the ordinary NCG method, becomes more simple. Using these constraints, we studied the renormalization group evolution of the Higgs boson mass. We adopted the idea of \'Alvarez et al. We determined the energy scale at which the NCG constraints hold by using one of the constraints, and then, by using another constraint, we obtained the initial condition to determine the RG evolution of the quartic self-coupling constant of the!
 Higgs field. The evolution indica
tes $m_H(\mu=m_H)=195 \mbox{~GeV}$.

Second, we studied that how we can obtain RGI constraints of the standard model in the NCG parametrization. We considered the cases that all coupling constants are expressed in terms of the $SU(2)$ gauge coupling $\alpha_{2}$. In particular, we focused on only $U(1)_{Y}$ and $SU(3)$ gauge couplings. The functional forms of $\alpha_{1}(\alpha_{2})$ and $\alpha_{3}(\alpha_{2})$ are determined uniquely from the RGI conditions and phenomenological data, so we determined the NCG parameters so as to give these functional forms.

We also considered the relation between the condition that a constraint among coupling constants of a model becomes RGI and the condition that the model becomes multiplicative renormalizable by using a Yukawa $+~\phi^{4}$ model. We showed that the condition of a RGI constraint is a necessary condition for the multiplicative renormalizability of the model whose bare coupling constants are constrained by the same constraint.

Constraints among coupling constants coming from the NCG method are derived from some artificial restrictions on the way of making the bosonic lagrangian from the field strengths defined in NCG, even if the restrictions have some naturalness. To obtain some constraints among coupling constants from the NCG method while ensuring the multiplicative renormalizability of the model, we should seek some symmetry that restricts the manner of constructing the lagrangian from the field strengths defined in NCG.
\section*{Acknowledgements}
The author wishes to express his sincere thanks to Professor S.~Naka for discussions. He is grateful to other members of his laboratory for their useful comments. He also thanks Professor I.~S.~Sogami and Professor K.~Morita for their encouragement and useful comments.

\end{document}